\documentclass[aps,prl,twocolumn,superscriptaddress,
               amsmath,floatfix]{revtex4}

\usepackage{graphicx}
\usepackage{dcolumn}
\usepackage{bm}
\usepackage{dcolumn}
\usepackage{subfigure}
\usepackage{color}
\usepackage{amsmath}
\usepackage{amssymb}
\usepackage{rotating}
\usepackage{wasysym}
\usepackage{ulem}
\usepackage{epstopdf}

\begin{document}

\title{\textbf{
Dy$_2$Ti$_2$O$_{7}$ Spin Ice: a Test  
Case for Emergent Clusters in a Frustrated Magnet\\
}}

\author{Taras Yavors'kii}
\affiliation{
Department of Physics and Astronomy, University of Waterloo, Ontario, N2L 3G1, Canada}
\author{Tom Fennell}
\affiliation{
London Centre for Nanotechnology, 17-19 Gordon Street, London, WC1H OAH, United Kingdom}
\author{Michel J.P. Gingras}
\affiliation{
Department of Physics and Astronomy, University of Waterloo, Ontario, N2L 3G1, Canada}
\affiliation{
Department of Physics and Astronomy, University of Canterbury, Private Bag 4800, Christchurch, New Zealand}
\author{Steven T. Bramwell}
\affiliation{
London Centre for Nanotechnology, 17-19 Gordon Street, London, WC1H OAH, United Kingdom}
\affiliation{
Department of Chemistry, University College of London, London, WC1H 0AJ, United Kingdom}

\date{\today}

\begin{abstract}
Dy$_2$Ti$_2$O$_7$ is a 
geometrically frustrated magnetic material
with a strongly correlated spin ice regime that extends from 1~K down
to as low as 60~mK. 
The diffuse elastic neutron scattering intensities
in the spin ice regime can be remarkably well described by a
phenomenological model of weakly interacting hexagonal spin clusters, as
invoked in other geometrically frustrated magnets.
We present a highly refined microscopic theory of Dy$_2$Ti$_2$O$_7$ 
that includes long range dipolar  and exchange interactions
to third nearest neighbors
and which demonstrates
that the clusters are purely fictitious in this material. 
The seeming emergence of composite spin clusters and their associated
scattering pattern is instead an indicator of 
fine-tuning of ancillary correlations within a
 strongly correlated state. 
\end{abstract}

\maketitle


Geometrically frustrated magnetic materials have in recent years furnished 
many new paradigms 
for the exploration of novel condensed states of matter.
Examples include spin ice behavior 
~\cite{Bramwell-Science,Harris97,Ramirez99,Higashinaka03,Hiroi03,
Sakakibara03,Tom-experiment,
Aoki04,Higashinaka05,Tabata06,Sato06}, 
collective paramagnetism or spin liquid behaviour~\cite{TbTiO}, 
spin glassiness in stoichiometrically pure materials~\cite{YMoO} 
and an anomalous Hall effect~\cite{taguchi_hall}. 
Frustration  is detrimental to the development
of conventional periodic arrangements of magnetic moments at low temperatures~\cite{FrNovel}.
As a consequence, at temperatures
that are small compared to the scale of the leading interactions, 
highly frustrated systems typically form collective  dynamical states~\cite{Villain}.
Most commonly, the massive degeneracy of such ``cooperative paramagnetic'' states
is ultimately resolved by specific material-dependent perturbative effects.
This is the origin of the diversity of experimentally observed phenomena~\cite{FrNovel}.

Among the many systems that exhibit geometric frustration \cite{FrNovel},
all the above phenomena have been observed in materials with magnetic moments (spins) 
residing on a network of corner sharing tetrahedra, or pyrochlore lattice.  
This arrangement is found, for example, in the rare earth titanates 
$R_2$Ti$_2$O$_7$, with magnetic rare earth ions ($R^{3+}$ = Ho, Dy $\dots$),  
and in the spinels like $AB_2$X$_4$ ($A$ = Zn, Cd), with magnetic 
transition metal ions ($B=$ Fe$^{3+}$, Cr$^{3+}$).  In this paper we
are concerned with Dy$_2$Ti$_2$O$_7$, a spin ice~\cite{Harris97,Bramwell-Science,Ramirez99}. 
In this material, the local spin correlations are characterized
by the ice rule: two spins point in and two 
spins point out of every tetrahedron~\cite{Harris97,Ramirez99, Bramwell-Science}.  
The strongly correlated spin ice regime (analogous to the spin liquid or collective paramagnetic
regime in geometrically frustrated antiferromagnets~\cite{Villain}) 
extends from 1 K down to the lowest measured temperatures ($\lesssim 100$ mK). 

Given the many phenomenologies exhibited by frustrated materials, the recent idea~\cite{LeeNature}
that there may be 
an organizing principle 
that describes the spin correlations in spin 
liquid states is very 
appealing.  
Using inelastic neutron scattering data, it was shown in Ref.~\cite{LeeNature}
that the spatial correlations of magnetic excitations in the spinel ZnCr$_2$O$_4$ 
are well described by a model of strongly bound hexagonal spin clusters that are 
uncorrelated with respect to each other.  
Subsequently, similar evidence of such clusters
was obtained 
in CdFe$_2$O$_4$~\cite{CdFe}, CdCr$_2$O$_4$~\cite{CdCrO} and
Y$_{0.5}$Ca$_{0.5}$BaCo$_4$O$_7$ \cite{Schweika07},
lending weight to the idea that the clusters may effectively be emergent 
objects.
In most cases it is
 numerically or analytically rather intractable
to examine these systems by detailed microscopic 
calculations. 
However,
Dy$_2$Ti$_2$O$_7$~\cite{Ramirez99} provides an exception, 
as we show here~\cite{static_vs_dynamics}. 

\begin{figure}[ht]
\includegraphics[width=3.4in,height=4.2in]{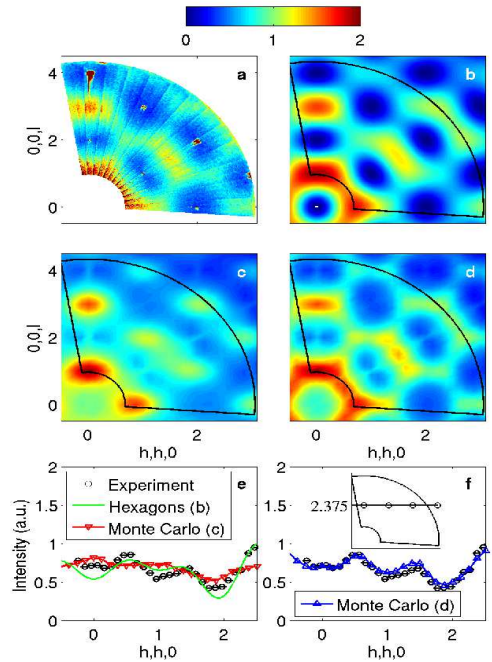}
\caption{(color online). 
Neutron scattering 
intensity $I({\bf q})$
 of Dy$_2$Ti$_2$O$_7$ as a function of the wavevector. 
{\bf (a)} Experimental 
elastic scattering intensity at 300~mK~\cite{Tom-experiment} in the (hhl) 
planes of the reciprocal space 
is notably amassed along the hexagonal
 zone boundaries. 
{\bf (b)} 
$I({\bf q})$
calculated using the model of uncorrelated hexagonal spin clusters (see text).
{\bf (c)} 
 $I({\bf q})$ obtained from 
MC simulations on the 
s-DSM of Dy$_2$Ti$_2$O$_7$ \cite{Hertog00}
describes the main experimental features, but is inadequate in 
reproducing the ZBS~\cite{Tom-experiment}.
{\bf (d)} The 
g-DSM allows for an excellent match between the theoretical 
and experimental $I({\bf q})$, and
allows us to identify
that correlations beyond 3rd nearest neighbors
are the microscopic origin 
behind the ZBS. 
{\bf (e,f)} 
Quantitative comparison of the experimental ({\bf a}) and theoretical ({\bf b}, {\bf c}, 
{\bf d}) data along a cut through the reciprocal space 
chosen to emphasize the mismatch with the s-DSM.
Panels {\bf c}, {\bf d}, {\bf e}, {\bf f} 
produced by MC simulations of $8192=8^3 \times 16$ spins.}
\label{Iqs.eps}
\end{figure}

In Ref.~\cite{Tom-experiment}, we and collaborators
showed that the well-established 
microscopic ``standard dipolar spin ice model''
(s-DSM, defined below~\cite{Bramwell-Science,Hertog00,Melko04})
describes the spin correlations in the spin ice state of 
Dy$_2$Ti$_2$O$_7$ with only modest success. 
The original objective for the present study 
was to understand the reason for the differences between theory and experiment. 
To do so, we  examined two models of the spin-spin correlations in Dy$_2$Ti$_2$O$_7$ 
by calculating the Fourier image, $I({\bf q})$, of the spin
correlation function and comparing it to the 
energy-integrated neutron scattering structure factor~\cite{Tom-experiment}.  
The Dy$^{3+}$ ion in Dy$_2$Ti$_2$O$_7$ 
has a Kramers doublet as  single-ion crystal field ground state~\cite{Rosenkranz}
that, for each site $i$, is well approximated by a classical Ising degree of 
freedom, $s_i=\pm 1$, 
defined along its 
 local $[111]$ trigonal axis
 ${\hat z}_i$~\cite{Hertog00}.
We calculate $I({\bf q})$ using~\cite{Tom-experiment}:
\begin{equation}
\label{Iq}
I({\bf q}) =
\frac{{[f( {\vert {\bf q} \vert} )]}^2} {N}
 \sum_{ij}
\langle {s}_{i} {s}_{j} \rangle \,
({\hat z}_{i}^{\perp}\cdot{\hat z}_{j}^{\perp})
e^{i {\bf {q}}\cdot{\bf {r}}_{ij}}\,,
\end{equation}
where $\langle \ldots \rangle$ 
denote thermally averaged $\langle s_i s_j\rangle$ 
correlations between the Ising spins at sites 
$i,\,j$; 
${\hat z}_i^\perp$ is the component of the quantization direction at site $i$ perpendicular
to the scattering vector ${\bf q}$, $N$ is the number of spins
and $f( {\vert {\bf q} \vert} )$ is the Dy$^{3+}$ magnetic form factor~\cite{Brown}.
For comparison with experiment, $I({\bf q})$ is adjusted by 
an overall scale factor and a slowly varying linear-in-$|{\bf q}|$ background.
The experimental diffuse neutron scattering, measured in the elastic approximation 
in the (hhl) plane
(from Ref.~\cite{Tom-experiment}), is shown in Fig.~\ref{Iqs.eps}a. 
As previously observed~\cite{Tom-experiment}, 
in addition to structural Bragg peaks (e.g. (004), (222))  
and bright
broad features at (001), (003) and ($\frac{3}{2}\frac{3}{2}\frac{3}{2}$), 
the experimental $I({\bf q})$ is further decorated by hexagonal 
loops of diffuse scattering running along the Brillouin zone boundaries. 

The first model of correlations we examine is phenomenological
and based on postulating an ansatz for
$\langle {s}_{i} {s}_{j} \rangle$ 
in (1).
We assume that $\langle {s}_{i} {s}_{j} \rangle$ can be viewed as generated
by static clusters that remain uncorrelated between themselves. Each cluster is 
a zero-magnetization hexagonal loop of spins that circulate perpendicular to the loop normal.
These clusters are the discrete 
equivalent of the ``emergent'' clusters used to describe the inelastic 
$I({\bf q})$ 
in ZnCr$_2$O$_4$~\cite{LeeNature} (see Fig.~1c in Ref.~\cite{LeeNature}).
Taking into account that all spins on a pyrochlore lattice can be grouped
 into non-overlapping hexagons without breaking the ice rules,  
that hexagon normals have four possible orientations, 
and that each hexagon has two possible senses of 
``spin circulation''
around the normal,
$I({\bf q})$ can then be calculated
 using Eq.~(1).
Fig.~\ref{Iqs.eps}b shows 
$I({\bf q})$
 calculated using the spin cluster scattering function. 
This model describes the experimental data quite well. 
The selection of 
hexagonal clusters as effective degrees of freedom in Dy$_2$Ti$_2$O$_7$
does not incorporate any microscopic information about the 
host material,
and thus be viewed as another example~\cite{LeeNature} 
 of emergent composite spin clusters 
in frustrated systems.

The second approach we use to determine $\langle s_i s_j \rangle$ is microscopic.
The s-DSM was previously shown to  account fairly well 
for the spin ice phenomenology of Dy$_2$Ti$_2$O$_7$ \cite{Hertog00,Bramwell-Science,Fukazawa02}. 
It comprises the  magneto-static dipole interaction, 
which gives a ferromagnetic nearest neighbor coupling, that competes
with a weaker antiferromagnetic nearest-neighbor exchange 
interaction $J_1$. 
Sufficiently strong 
antiferromagnetic $J_1$ would lead to long range order \cite{Hertog00,Melko04}. 
It is the 
 interplay between the properties of the spin ice manifold 
and the symmetry and long range nature 
of the dipolar interaction 
that leads to a correlated spin ice 
state over an extended temperature range~\cite{Hertog00,Melko04,Isakov05}.
Surprisingly, 
the s-DSM is much less successful than the simple phenomenological cluster 
model at describing 
$I({\bf q})$.

Monte Carlo 
(MC) simulations have shown~\cite{Tom-experiment} 
that the s-DSM correctly describes the location and relative intensity of the strong 
$I({\bf q})$ features, 
but fails to reproduce the hexagonal zone 
boundary scattering (ZBS), Fig.~\ref{Iqs.eps}c.
We interpret this as a sign that the
s-DSM is incomplete and needs to be extended~\cite{Jacob-PRL,Tabata06}.

Dy$_2$Ti$_2$O$_7$ displays a number of phase transitions and other structured response 
in an applied magnetic field 
${\bf H}$~\cite{Ramirez99,Higashinaka03,Hiroi03,Sakakibara03,
Aoki04,Higashinaka05,Sato06,Tabata06}. 
Independently of the disagreement between Fig.~\ref{Iqs.eps}a and 
Fig.~\ref{Iqs.eps}c, the necessity to
adjust the s-DSM was previously
 also suggested by the observation that,
 while it qualitatively explains
the in-field transitions,
the s-DSM does not correctly predict their temperatures~\cite{Jacob-PRL}, 
unless properly adjusted by
perturbative exchange couplings beyond $J_1$.
Here, we consider a generalized dipolar spin ice model (g-DSM) of Dy$_2$Ti$_2$O$_7$ that 
includes second $J_2$ and third $J_3$ nearest neighbor exchange couplings:
\begin{eqnarray}
\nonumber
&&{\cal H}=\sum_{i>j} s_i s_j\,\,
\left\{ \rule{0pt}{18pt}
\; \sum_{\nu=1}^3 \; J_{\nu} \;\; \delta_{r_{ij},r_{\nu}}\; {\hat z}_i\cdot{\hat z}_j\; +
D{r_{1}^3}/{r_{ij}^3}\, \left[ {\hat z}_i\cdot{\hat z}_j
\right. \right. 
\\
&&
\left.\left.
-3\,({\hat z}_i\cdot\hat{r}_{ij}) ({\hat z}_j
\cdot\hat{r}_{ij})\right]
\rule{0pt}{18pt} \right\} - g \langle\! J^z \!\rangle \mu_{\rm B} \sum_{i} s_i ({\hat z}_i\cdot\bf{H})\,.
\label{Ham}
\end{eqnarray}
Here, $i$, $j$ span the sites of the $\rm Dy^{3+}$ ions,
$r_{ij}$ and $\hat{r}_{ij}$ is the length and direction of
the vector separation between spin pairs. 
$D=\mu_0  (\langle \! J^z \! \rangle g \mu_{\rm B} )^2 /(4\pi {r_{1}}^3)=
1.3224 $ K is the strength of the dipolar interaction at  nearest-neighbor distance  
$r_{1}=a\sqrt{2}/4$, as obtained from the estimate
$\langle \! J^z \!\rangle  = 7.40$ \cite{Tb-moment} within the ground state doublet of 
Dy$^{3+}$ in Dy$_2$Ti$_2$O$_7$, and from the size 
$a=10.124$ \AA ~\cite{Fukazawa02}
of the cubic unit cell of the material. $g=4/3$ is the Dy$^{3+}$ Land\'e factor, 
$\mu_{\rm B}$ is 
the Bohr magneton, and
$J_\nu$ is the exchange coupling
 of spins at distance $r_\nu$. 
In reality, there should be two different 3rd nearest 
neighbor exchange couplings.
We take them equal, which, as shown below,
provides a consistent description of Dy$_2$Ti$_2$O$_7$.
By doing mean-field theory (MFT) calculations of 
$I({\bf q})$ in the paramagnetic regime~\cite{Enjalran}
 we checked that the conclusions about  $I({\bf q})$
are not sensitive to $\sim 50$\% deviations of the two $J_3$ from equality.

\begin{figure}[t]
\includegraphics[angle=-90,width=3.5in]{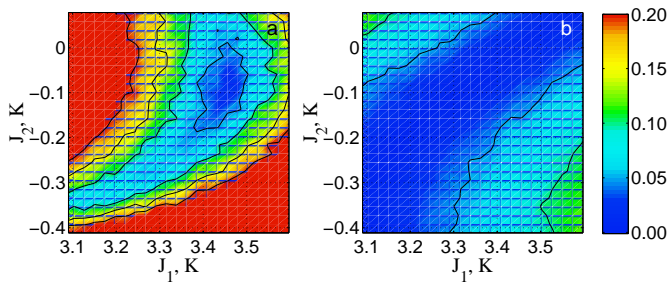}
\caption{(color online). 
rms deviation $\sigma$ of Monte Carlo vs. experimental~\cite{Higashinaka03} specific heat data
in the $J_1-J_2$ plane at fixed value of $J_3$ = 0.025~K. 
The isolines are drawn at $\sigma = 0.05, 0.10, 0.15$ and $0.20$ 
J mol$^{-1}$K$^{-2}$ outward. 
The two functions $\sigma$ [(a), obtained 
for the $T_1$=0.7~K-1.49~K temperature interval) 
and (b), for $T_2$=1.5~K-2.8~K], 
yield candidate solutions (delineated by inner isolines)
in overlapping parts of the $J_1-J_2$ plane. 
The determination of the $J_\nu$ couplings in the Hamiltonian (~\ref{Ham})
of Dy$_2$Ti$_2$O$_7$  is obtained  via a fitting 
to several bulk  experimental thermodynamic data, akin to the procedure illustrated here
(see text).}
\label{Cv_fit.eps}
\end{figure}

Rather extensive 
MC simulations were used to investigate whether the g-DSM can provide 
a consistent description of the phenomenology of Dy$_2$Ti$_2$O$_7$ both 
for ${\bf H}=0$ \cite{Ramirez99,Hiroi03,Higashinaka03} and ${\bf H}\neq 0$ 
\cite{Ramirez99,Sakakibara03,Aoki04,Higashinaka05,Sato06} regimes.
MC simulations were  performed on system sizes of 
$1024=4^3\times 16$ spins with periodic boundary conditions, treating 
the long range dipolar interaction by the Ewald method~\cite{Melko04}. 
For each set of parameters in (\ref{Ham}) and each temperature,
$10^5$ MC steps per spin for equilibration followed by an 
additional $10^6$ steps per spin for production
were performed.
The simulations exploited both 
single spin-flip and 
loop-update~\cite{Melko04}
 Metropolis algorithms,
the latter being 
a generalized algorithmic MC version of the dynamics of the hexagonal cluster degrees 
of freedom discussed above 
(cf. Fig.~1 \cite{Melko04}).
 Finite size effects were found to be 
insignificant for the discussion below.

Three groups of experimental data about
Dy$_2$Ti$_2$O$_7$ are used to determine 
the $J_\nu$   in  (\ref{Ham}).
First, we consider the temperature-driven 
ferromagnetic ordering of Dy$_2$Ti$_2$O$_7$ 
for ${\bf H}$ nearly along the [112] direction \cite{Higashinaka05,Jacob-PRL}. 
For fixed $D$, the transition is 
controlled by $J_3$ only \cite{Jacob-PRL}. 
Mapping the MC critical temperature, 
obtained as location of the
 maximum in magnetic heat capacity $C_m$, to the experimental
estimates of 0.34, 0.28(1) and 0.26(1)~K, obtained by specific heat~\cite{Ramirez99}, 
susceptibility~\cite{Higashinaka05} and magnetization~\cite{Sato06} measurements,
we conclude that 0.019~K $\apprle \! J_3  \! \apprle$ 0.026~K.
Second, we examine $C_m(T)$ data 
in ${\bf H}=0$
\cite{Ramirez99,Higashinaka03,Hiroi03}
for constraining $J_1$, $J_2$, as $C_m(T)$ is 
only weakly sensitive to the allowed small variation of $J_3$.
We determine optimal $J_1$, $J_2$ by minimizing 
the root mean square (RMS) deviation $\sigma$ 
of experimental vs theoretical temperature curves for 
$C_m(T)/T$.
The function $\sigma(J_1,J_2)$ at $J_3=0.025$~K is plotted in Fig.~\ref{Cv_fit.eps}
using Ref.~\cite{Higashinaka03} data.
The two panels illustrate that $\sigma(J_1,J_2)$ has a stable minimum even if 
determined over two different temperature intervals.
Third, we exploit an empirical equation 
${\rm H}_c=0.90(1)+0.08T,\,T\apprle 0.36$~K~\cite{Sakakibara03,Aoki04} 
for the line of phase transitions in Dy$_2$Ti$_2$O$_7$ for $\bf{H}$ along [111].
We obtain values of ${\rm H}_c$ at several temperatures as positions of maxima 
on MC $C_m(\bf{H})$ curves,
and vary $J_1$ and $J_2$
to match the experimental ${\rm H}_c(T)$.
This, the above ${\bf H}=0$ 
 analysis and an additional observation 
that at ${\bf H}=0$ a magnetic phase transition, if any, can occur only at 
$T_c<300$~mK~\cite{Bramwell-Science,Tom-experiment}, 
cf. Fig.~\ref{Iqs.eps}a, yields:
  3.26~K $\apprle \! J_1  \! \apprle$  3.53~K;
$- 0.20$ ~K $\apprle \! J_2  \! <$ 0~K.
Our value for $J_2$ is consistent with the value 
$J_2=-0.1$ K reported in Ref.~\cite{Tabata06} from fitting 
(\ref{Ham}) 
with $D=1.41$~K, $J_1=3.72$~K, $J_3=0.03$~K 
(Ref.~[\onlinecite{Jacob-PRL}])
 to the
experimental $C_m({\bf H})$ for ${\bf H}$ along [111]. 

To verify whether the experimental $I({\bf q})$ in  Fig.~\ref{Iqs.eps}a 
can be described by (\ref{Ham}), we perform MC 
simulations of the optimized g-DSM  at 300~mK. 
We find that the g-DSM with $J_\nu$ 
couplings within the above allowed limits is consistent 
with the experimental scattering pattern in Fig.~\ref{Iqs.eps}a,
but requires even stronger restriction on $J_2$: 
$- 0.16$ ~K $\apprle \! J_2  \! \apprle$ $-0.10$ ~K.
For instance, the theoretical pattern at $J_1=3.41\,{\rm K},\, 
J_2=-0.14\,{\rm K},\, J_3=0.025\,{\rm K}$ (Figs.~\ref{Iqs.eps}d, f)  
reproduces the experimental one (Fig.~\ref{Iqs.eps}a) extremely well. 
As seen from 
Figs.~\ref{Iqs.eps}c, d, e and f,
the small adjustments to the initial 
s-DSM \cite{Hertog00} 
 do result in a redistribution of the scattering response, 
with intensities that now capture correctly both major features of the experimental pattern, 
 and the weaker ZBS. 
The characteristic features on the experimental $I({\bf q})$ for Dy$_2$Ti$_2$O$_7$, 
first described above as arising
 from hexagonal spin clusters,
can therefore be ``straightforwardly'' described within a conventional microscopic
 treatment of the static spin-spin correlation function~\cite{footnote}.

Having now a credible microscopic model in hand, 
are we able to explain the success of the phenomenological fit (Fig.~\ref{Iqs.eps}b)?
To answer this question,
we examine the direct space $\langle s_i s_j\rangle$ 
 correlations behind
the reciprocal space patterns in Fig.~\ref{Iqs.eps}b, c, d. 
We intimate that
the picture of independent clusters
is equivalent to a 
correlation function truncated beyond third nearest neighbor distance
(i.e. outside the hexagonal cluster).
The $\langle s_i s_j\rangle$
obtained from MC simulations on both the 
s-DSM (Fig.~\ref{Iqs.eps}c) and optimized g-DSM (Fig.~\ref{Iqs.eps}d)
are very similar at short distances.
If truncated beyond third nearest neighbor distance, 
the Fourier transform of each
produces a pattern close to that of experiment
(Fig.~\ref{Iqs.eps}a).
We conclude that the difference between the two MC $I({\bf q})$ patterns
(Fig.~\ref{Iqs.eps}c and Fig.~\ref{Iqs.eps}d)
is caused by 
correlations 
{\it beyond} third neighbors.
For the tuned g-DSM, 
 the long range part strongly 
reinforces the ZBS arising from the short range part, 
contributing the {\it majority} of this scattering.
This contradicts the notion of independent 
hexagons, which underly 
the calculation of $I({\bf q})$ in Fig.~\ref{Iqs.eps}b,
since correlations between
hexagons are of the same order as the correlations defining the
hexagons.
Thus, weakly interacting clusters are not an appropriate microscopic 
description of Dy$_2$Ti$_2$O$_7$, nor
are they here an effective organizing principle.
Rather, the ZBS are caused
 by subsidiary ($J_2$ and $J_3$) interactions that are largely 
inconsequential to the formation of the strongly correlated 
spin ice regime~\cite{Hertog00,Melko04,Isakov05}. 
We further confirmed the fine-tuned nature of the ZBS
using MFT to calculate 
$I({\bf q})$ in the paramagnetic regime~\cite{Enjalran}.
Finally, we 
 note that independent hexagonal clusters,
unlike the s-DSM and g-DSM,
fail to predict the ``pinch point'' scattering
characteristic of
 a spin ice manifold, which is
discernable in the experiment.

We finally comment on the broader consequences that the observed complementarity 
of the phenomenological and microscopic interpretations of neutron scattering 
patterns might have for other systems. Our results support the
 idea expressed in Ref.~\cite{LeeNature} that 
cluster-like scattering (CLS)
may be common among highly frustrated magnets and establish here
a microscopic mechanism by which it can arise.
However, 
by considering Dy$_2$Ti$_2$O$_7$ as a test case,
we have shown that 
the CLS does  not necessarily imply the
emergence of ``real''
clusters~\cite{Henley05},
 nor a new organizing principle. 
Instead, the CLS is the property of a strongly correlated 
liquid-like state
 and a consequence of the sensitivity of frustrated 
systems to perturbations~\cite{Villain}. 
Our study also shows how neutron scattering can 
allow one to fully interpret CLS
provided sufficiently accurate data and a computationally
tractable theoretical microscopic model are
 available. 


We thank our collaborators in Ref.~\cite{Tom-experiment},
particularly B. F{\aa}k and O. Petrenko, for their role in 
obtaining the neutron data.
We acknowledge useful discussions with 
C. Broholm, M. Enjalran, 
C. Henley, S.-H. Lee, H. Molavian, O. Tchernyshyov, and A.-M. Tremblay.
Support for this work was provided by NSERC,
the CRC Program (Tier I, M.G), the CFI,
the OIT, and the CIFAR.
M.G. acknowledges the U. of Canterbury (UC) for financial support
and the hospitality of the Dept. of Physics and Astronomy at UC.


\vspace{-4mm}
\begin{thebibliography}{99}
\vspace{-4mm}

\bibitem{Harris97} M. J. Harris {\it et al.}, Phys. Rev. Lett. {\bf 79}, 2554
(1997).


\bibitem{Ramirez99}
A. P. Ramirez {\it et al.}, Nature {\bf 399}, 333 (1999).

\bibitem{Bramwell-Science}
S. T. Bramwell and M. J. P. Gingras,
Science {\bf 294}, 1495 (2001).


\bibitem{Higashinaka03}
R.~Higashinaka and Y.~Maeno,
Phys. Rev. B {\bf 68}, 014415 (2003).

\bibitem{Hiroi03}
Z.~Hiroi {\it et al.}, Phys. Soc. Jpn. {\bf 72}, 411 (2003).

\bibitem{Sakakibara03}
T.~Sakakibara {\it et al.}, Phys. Rev. Lett. {\bf 90}, 207205 (2003).


\bibitem{Tom-experiment}
T. Fennell {\it et al.}, Phys. Rev. B {\bf 70}, 134408 (2004).


\bibitem{Aoki04}
H.~Aoki {\it et al.}, Phys. Soc. Jpn. {\bf 73}, 2851 (2004).


\bibitem{Higashinaka05}
R.~Higashinaka and Y.~Maeno,
Phys. Rev. Lett. {\bf 95}, 237208 (2005).

\bibitem{Sato06}
H.~Sato {\it et al.},
J. Phys.: Condens. Matter {\bf 18}, L297 (2006).

\bibitem{Tabata06}
Y.~Tabata {\it et al.}, Phys. Rev. Lett. {\bf 97}, 257205 (2006).

\bibitem{TbTiO}
J. S. Gardner {\it et al.}, Phys. Rev. Lett. {\bf 82}, 1012 (1999).

\bibitem{YMoO}
B.~D.~Gaulin {\it et al.}, Phys. Rev. Lett. {\bf 69}, 3244 (1992).

\bibitem{taguchi_hall} Y. Taguchi {\it et al.},  Science {\bf 291,} 2573
(2001).

\bibitem{FrNovel}
{\it Frustrated Spin Systems}, ed.~H.T.~Diep, World Scientific,
Singapore (2004);
A. P. Ramirez, in
{\it Handbook of Magnetic Materials}, ed. K. H. J. Buschow (Elsevier Science, Amsterdam, 2001), Vol. 13;
J.E. Greedan,
J. of Alloys and Compounds {\bf 408-412}, 444 (2006).

\bibitem{Villain}
J. Villain, Z. Phys. B {\bf 33}, 31 (1979).

\bibitem{LeeNature}
S.-H. Lee {\it et al.}, Nature {\bf 418}, 856 (2002). 


\bibitem{CdFe}K. Kamazawa {\it et al.}, Phys. Rev. B {\bf 70}, 024418 (2004).

\bibitem{CdCrO}
J.-H. Chung {\it et al.}, Phys. Rev. Lett. {\bf 95}, 247204 (2005).

\bibitem{Schweika07}
W. Schweika {\it et al.}, 
Phys. Rev. Lett. {\bf 98}, 067201 (2007).

\bibitem{static_vs_dynamics} 
Unlike Ref.~\cite{LeeNature}, we base our study 
on elastic, rather than inelastic, neutron scattering data, Ref.~\cite{Tom-experiment}.


\bibitem{Hertog00}
B. C. den Hertog and M. J. P. Gingras,
Phys. Rev. Lett. {\bf 84}, 3430 (2000).


\bibitem{Melko04}
R. G. Melko and M. J. P. Gingras,
J. Phys.: Condens. Matter {\bf 16}, R1277 (2004).



\bibitem{Rosenkranz}
S. Rosenkranz {\it et al.}, J. Appl. Phys. {\bf 87}, 5914 (2000).

\bibitem{Brown} P.~J.~Brown, Magnetic form factors, Chapter 4.4.5, in:
{\it International Tables for Crystallography}, vol.~C, pp. 391-399,
edited by A.~J.~C.~Wilson (Dordrecht, Holland: D. Reidel Pub. Co., 1983-1993).


\bibitem{Fukazawa02}
H. ~Fukazawa {\it et al.}, Phys. Rev. B {\bf 65}, 054410 (2002).


\bibitem{Isakov05}
S. V. Isakov {\it et al.},
Phys. Rev. Lett. {\bf 95}, 217201 (2005).


\bibitem{Jacob-PRL}
J.~P.~C.~Ruff {\it et al.},
Phys. Rev. Lett. {\bf 95}, 097202 (2005).

\bibitem{Tb-moment}
The Dy$^{3+}$ dipole moment was obtained by diagonalizing the
crystal-field (CF) Hamiltonian for Dy$_2$Ti$_2$O$_7$,  using the
CF parameters taken from Ref.~\cite{Rosenkranz} 
for Ho$_2$Ti$_2$O$_7$ but rescaled for Dy$_2$Ti$_2$O$_7$.


\bibitem{Enjalran}
M. Enjalran and M. J. P. Gingras, Phys. Rev. B {\bf 70}, 174426 (2004).

\bibitem{footnote}
Examining models (\ref{Ham}) with couplings $J_{\nu}$ close to those used in 
Fig.~\ref{Iqs.eps}d, we find evidence that the model exhibits a 
transition to long-range ordered state at a temperature as low as 
60~mK, unlike 180~mK found in Ref.~\cite{Melko04}.

\bibitem{Henley05}
A similar view is expressed in C.~Henley, Phys. Rev. B {\bf 71}, 014424 (2005).



\end{thebibliography}
\end{document}